\documentclass[aps,prl,showpacs,superscriptaddress,twocolumn]{revtex4}

\usepackage{amsmath}
\usepackage{graphicx}
\usepackage{longtable}
\usepackage[ansinew]{inputenc}
\usepackage{color,ulem}
\usepackage{SIunits}

\newcommand{\BiSe}{Bi$_2$Se$_3$}

\newcommand{\beq}{\begin{eqnarray}}
\newcommand{\eeq}{\end{eqnarray}}

\newcommand {\tw}[1]{{\color{black} #1}}
\newcommand {\ak}[1]{{\color{black} #1}}
\newcommand {\jw}[1]{{\color{black} #1}}

\begin{document}


\title{Controllable magnetic doping of the surface state of a topological insulator}

\author{T. Schlenk}
\affiliation{Institute for Applied Physics, Universit\"{a}t Hamburg, D-20355 Hamburg, Germany}

\author{M. Bianchi}
\affiliation{Department of Physics and Astronomy, Interdisciplinary Nanoscience Center, Aarhus University, 8000 Aarhus C, Denmark}

\author{M. Koleini}
\affiliation{Bremen Center for Computational Materials Science, University of Bremen, D-28359 Bremen, Germany}

\author{A. Eich}
\affiliation{Institute for Applied Physics, Universit\"{a}t Hamburg, D-20355 Hamburg, Germany}

\author{O. Pietzsch}
\affiliation{Institute for Applied Physics, Universit\"{a}t Hamburg, D-20355 Hamburg, Germany}

\author{T. O. Wehling}
\affiliation{Bremen Center for Computational Materials Science, University of Bremen, D-28359 Bremen, Germany}
\affiliation{Institute for Theoretical Physics, Bremen University, D-28359 Bremen, Germany}

\author{T. Frauenheim}
\affiliation{Bremen Center for Computational Materials Science, University of Bremen, D-28359 Bremen, Germany}

\author{A. Balatsky}
\affiliation{Theoretical Division and Center for Integrated Nanotechnologies, Los Alamos National Laboratory, Los Alamos, New Mexico 87545, USA}
\affiliation{Nordic Institute for Theoretical Physics (NORDITA), S-106 91 Stockholm, Sweden}

\author{J.-L. Mi}
\affiliation{Center for Materials Crystallography, Department of Chemistry, Interdisciplinary Nanoscience Center, Aarhus University, 8000 Aarhus C, Denmark}

\author{B. B. Iversen}
\affiliation{Center for Materials Crystallography, Department of Chemistry, Interdisciplinary Nanoscience Center, Aarhus University, 8000 Aarhus C, Denmark}

\author{J. Wiebe}
 \email[Email: ]{jwiebe@physnet.uni-hamburg.de}
\affiliation{Institute for Applied Physics, Universit\"{a}t Hamburg, D-20355 Hamburg, Germany}

\author{A. A. Khajetoorians}
 \email[Corresponding author: ]{akhajeto@physnet.uni-hamburg.de}
\affiliation{Institute for Applied Physics, Universit\"{a}t Hamburg, D-20355 Hamburg, Germany}

\author{Ph. Hofmann}
\affiliation{Department of Physics and Astronomy, Interdisciplinary Nanoscience Center, Aarhus University, 8000 Aarhus C, Denmark}

\author{R. Wiesendanger}
\affiliation{Institute for Applied Physics, Universit\"{a}t Hamburg, D-20355 Hamburg, Germany}

\date{\today}

\begin{abstract}
\noindent A combined experimental and theoretical study of doping individual Fe atoms into \ak{\BiSe~is presented}. It is shown through a scanning tunneling microscopy study that single Fe atoms initially located at hollow sites \ak{on top} of the surface (adatoms) can be incorporated into subsurface layers by thermally-activated diffusion. Angle-resolved photoemission spectroscopy in combination with \textit{ab-initio} calculations suggest that the doping behavior changes from electron donation for the Fe adatom to neutral or electron acceptance for Fe incorporated into substitutional Bi sites.  According to \ak{first principles} calculations \ak{within density functional theory,} these Fe substitutional impurities retain a large magnetic moment thus presenting an alternative scheme for magnetically doping the topological surface state.  For both types of Fe doping, we see no indication of a gap at the Dirac point.
\end{abstract}



\maketitle%

Three-dimensional topological insulators (3D TIs) have a gapped bulk band structure but a spin-orbit interaction induced band inversion leads to the existence of a metallic topological surface state (TSS). The TSS can exhibit a Dirac-like dispersion which is protected by time-reversal symmetry, resulting in unique properties such as spin-momentum locking and prohibited backscattering \cite{Hasan2010,Qi2011}.  Nevertheless, defect scattering is possible for a two-dimensional TSS and scattering by magnetic impurities is of particular interest, as these can break time-reversal symmetry in the system. This may affect the topological protection and could potentially open a gap in the topological protected band connection at the surface \cite{Liu2009,Biswas2010,Chen2010,Wray2011}.

Two experimental approaches have been utilized to introduce magnetic impurities in 3D TIs, namely bulk magnetic doping  in crystals or thin films ~\cite{Liu2012,Xu2012,Chen2010,Hor2010,Vobornik2011}, and \textit{in-situ} surface doping via adsorption of impurities on pristine topological insulators~\cite{Wray2011,Valla2012,Scholz2012}. Surface doping can be easier to control experimentally and has the additional advantage that the bulk material remains a TI, something that is not necessarily the case for too high concentrations of bulk dopants~\cite{Liu2012}. Angle-resolved photoemission spectroscopy (ARPES) measurements of magnetically-doped Bi-chalcogenide samples prepared in both manners exhibit a feature near the DP which resembles a gap~\cite{Chen2010,Wray2011,Xu2012}. For the surface-doped case, however, it turned out that similar features can be generated by doping with non-magnetic impurities ~\cite{Bianchi2011}. Indeed, an apparent gap opening can be caused via the adsorbate-induced strong band bending near the surface~\cite{Bianchi2010,King2011}. 
  \ak{More recently, gaps in bulk magnetically doped thin \BiSe~films have been observed~\cite{Xu2012} resulting in a spin reorientation of the TSS.}  Finally, it has been shown that potential disorder induced by charged bulk defects hinders a clear analysis of faint effects due to the breaking of time-reversal symmetry on the scattering of the TSS electrons~\cite{Beidenkopf2011}. 
Therefore, a method where \textit{neutral} magnetic impurities can be introduced into the surface and magnetically interact with the TSS\ak{, \textit{while only weakly perturbing the band structure,}} is highly desirable.

\begin{figure*}[tbp]
\includegraphics[width=0.7\textwidth]{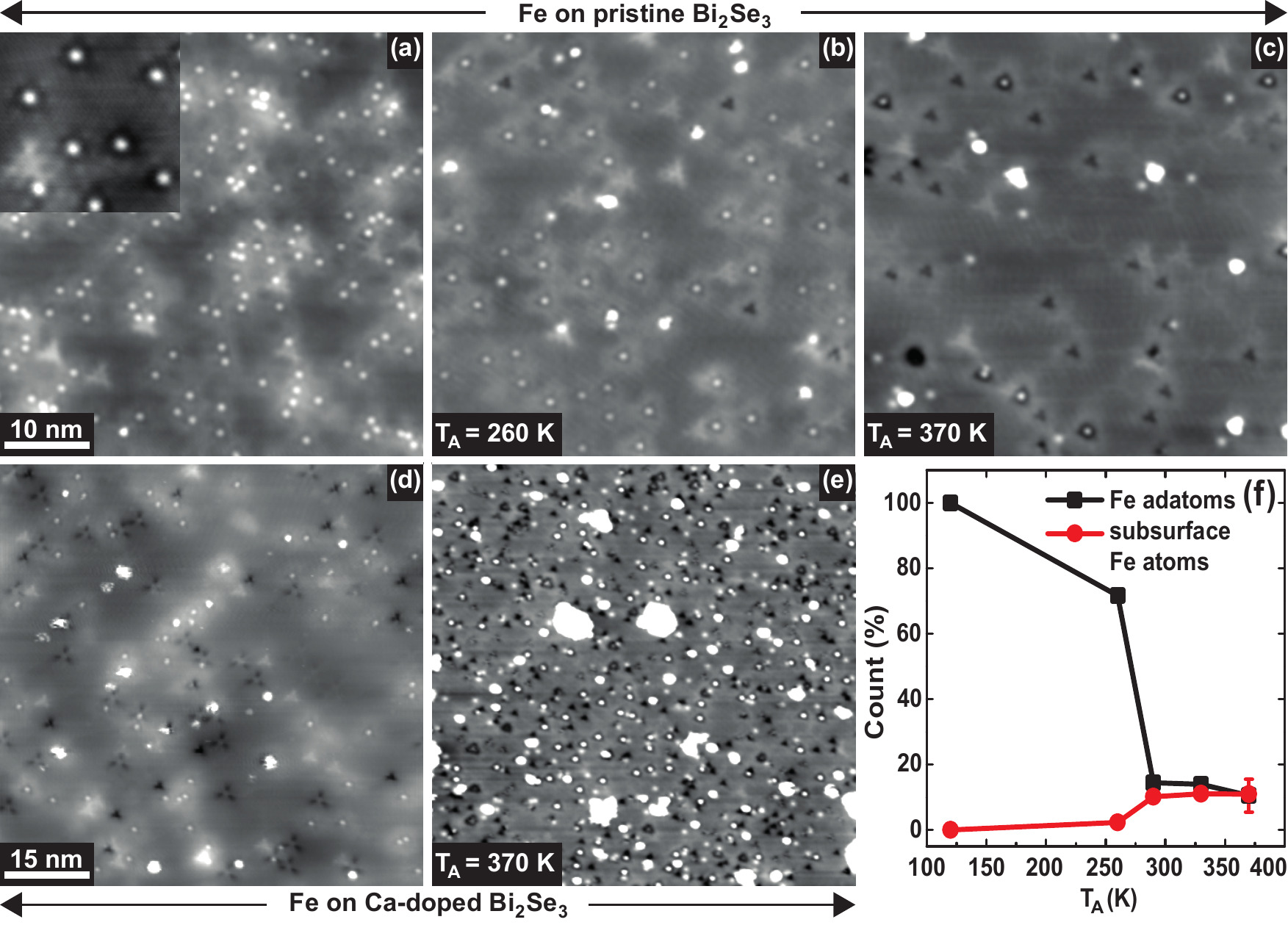}
\caption{\label{STM} Constant current STM images of differently prepared samples of Fe on \BiSe~($I_{\rm t}=50$pA, $V_{\rm b}=0.25$V (a-d), $V_{\rm b}=1$V (e).) (a) Cold deposited Fe ($\approx$ 1\% of a ML) on pristine \BiSe. Inset: magnified view showing 8 individual Fe adatoms ($10$nm $\times$ $10$nm, $V_{\rm b}=-0.3$V, $I_{\rm t}=1$nA). (b), (c) Samples prepared as in (a) annealed to $T_{\rm A} \approx 260$~K and $370$~K, respectively. (d) Cold deposited Fe ($0.5$\% ML) on Ca-doped \BiSe. (e) Sample prepared as in (d) but with $6$\% ML Fe and annealed to $T_{\rm A} \approx 370$~K. (f) Number of Fe adatoms and subsurface Fe atoms relative to initial Fe adatom number as a function of $T_{\rm{A}}$. \ak{Emergent clusters were not counted in the statistics.}}
\end{figure*}

In this letter, we demonstrate a way to control the charge state of a magnetic impurity such that this is achieved. Scanning tunneling microscopy (STM) experiments illustrate that Fe atoms that are adsorbed on the surface (adatoms) diffuse into subsurface sites of \BiSe~upon thermal annealing. Such bulk doping via the surface can be realized for a large range of coverages. Complementary ARPES studies \ak{illustrate} that annealing reduces the downwards band-bending, \ak{initially} produced by Fe adatoms, and restores the DP near its \ak{initial} location before Fe adsorption. These effects are seen regardless of the initial location of the DP, namely for both pristine (\textit{n}-doped) and Ca-doped (nearly intrinsic) substrates. The difference in doping behavior of Fe adatoms compared to incorporated subsurface Fe atoms at different lattice sites is further studied by first principles calculations \ak{within density functional theory}. The calculations reveal that the doping character changes from electron donor for Fe adatoms towards neutral for Fe \ak{incorporated into} Bi substitutional sites consistent with the ARPES results. \ak{Such} incorporated Fe atoms theoretically retain a large magnetic moment which is a prerequisite for \ak{modifying} the TSS by breaking the time-reversal symmetry locally.

STM experiments were carried out in a multichamber UHV-system with a base pressure below $1 \times 10^{-10}$~mbar utilizing a home-built variable temperature scanning tunneling microscope (VT-STM) \cite{Eremeev2012}. Annealed tungsten tips were used for all experiments and both tip and sample were cooled to $T=30$~K. STM topography images were taken in constant current mode at a tunneling current $I_{\rm t}$ with the bias voltage $V_{\rm b}$ applied to the sample.
ARPES measurements were performed at the SGM-3 beamline of the ASTRID synchrotron radiation facility \cite{Hoffmann2004} at a sample temperature of $T=60$~K. The angular and combined energy resolutions were $0.13^{\circ}$ and better than $15$~meV, respectively. Spectra were obtained along the $\bar{K}-\bar{\Gamma}-\bar{K}$ azimuthal direction with a photon energy of $h \nu =19.2$~eV.
Pristine and Ca-doped \BiSe~single crystals were grown as described in refs.~\cite{Bianchi2010,Bianchi2012}. In both experimental setups the \BiSe~samples were cleaved \textit{in situ} at room temperature (RT) and immediately cooled to temperatures of $T \leq 150$~K. 
Fe was \jw{originally} deposited onto the cold surface by e-beam evaporation resulting in a distribution of single Fe adatoms \cite{Honolka2012}.

\tw{To achieve a theoretical description of the Fe/Bi$_2$Se$_3$ system, we have performed density functional theory calculations with the VASP code~\cite{PhysRevB.47.558,PhysRevB.54.11169}, where we employed the GGA-PBE approximation to the exchange and correlation functional~\cite{PhysRevLett.77.3865} and used plane wave basis sets with a kinetic energy cutoff of 20~Ry in combination with \jw{projector} augmented-wave (PAW) pseudopotentials~\cite{PhysRevB.50.17953,PhysRevB.59.1758}. We used a ($3\times3$) supercell with two quintuple slabs which have been relaxed until the maximum force component of every atom was lower than 0.02~eV/\AA. A ($6\times6$) {\it k}-point grid \jw{has} been employed to obtain \jw{the} precise density of states \jw{(DOS)}.}

\begin{figure}[tbp]
\includegraphics[width=\columnwidth]{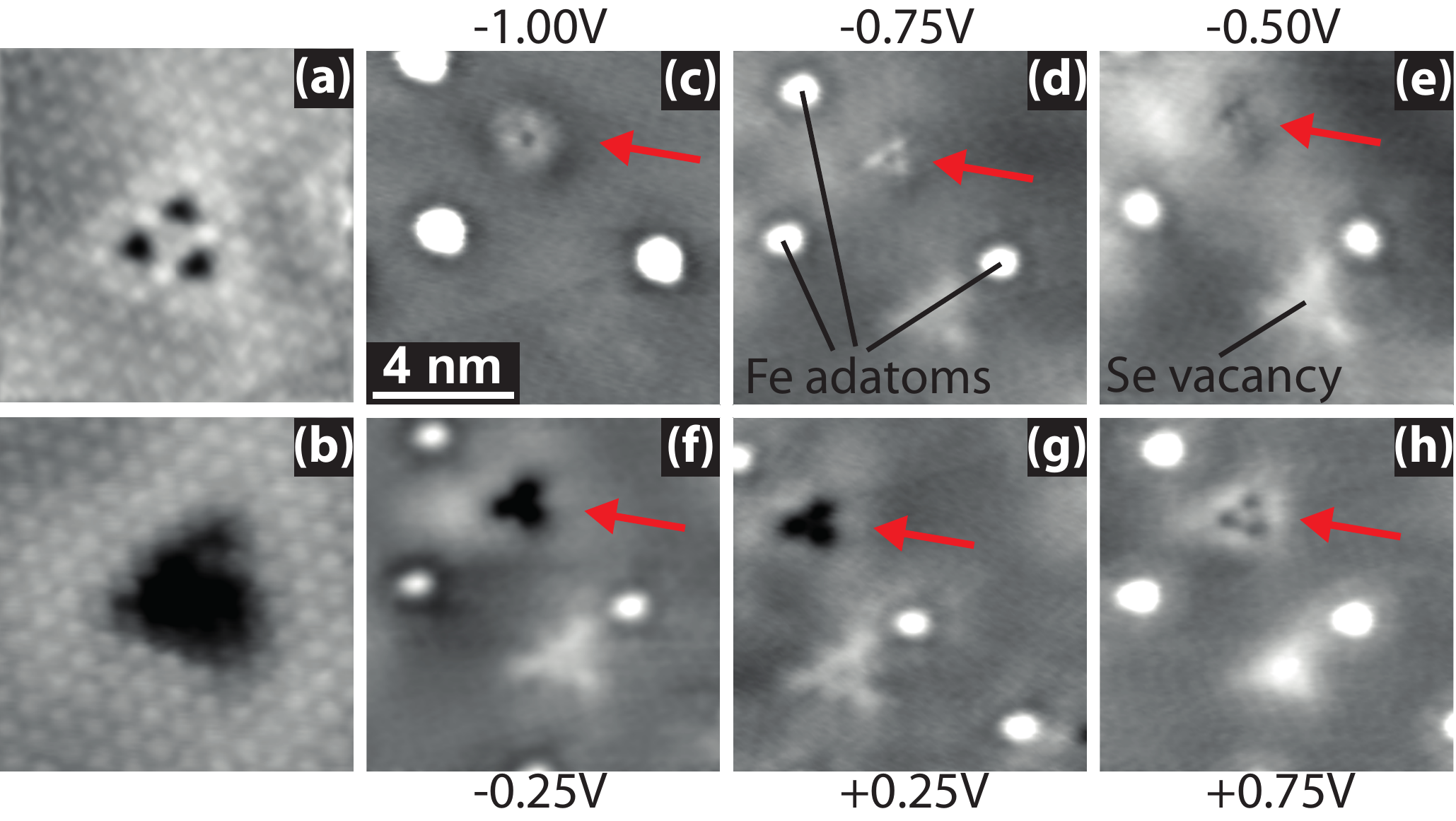}
\caption{\label{STM2} (a), (b) Atomically resolved STM images of two different subsurface Fe defects [$I_{\rm t}=50$pA, $V_{\rm b}=1$V (a), $V_{\rm b}=0.25$V (b)]. (c)-(h) Bias dependent constant current images of a subsurface Fe atom (red arrows), 3 Fe adatoms and a Se vacancy at $V_{\rm b}$ as indicated. ($I_{\rm t}=100$pA, $\Delta$\textit{z} = $35$pm).}
\end{figure}

Fig.~\ref{STM} shows a sequence of STM images where Fe was initially deposited onto the cold surface ($T \leq 150$~K) of pristine  (a-c) and Ca-doped (d,e) \BiSe. After deposition, the sample was annealed to a given temperature $T_{\rm A}$ as indicated in the image. For each subsequent annealing temperature, the sample was re-cleaved and the process repeated.  After cold deposition (Fig.~\ref{STM}(a) and inset), STM reveals that the surface is covered primarily by individual Fe adatoms which are strongly relaxed into either of the two hollow sites of the hexagonal lattice of Se surface atoms like seen in ref.~\cite{Honolka2012}. After annealing to $T_{\rm A} \approx 260$~K (Fig.~\ref{STM}(b)), a new type of defect appears, which is imaged as a triangular-shaped depression at $V_{\rm b}=0.25$~V. Simultaneously, a decrease in the overall density of Fe adatoms and the formation of clusters are observed (larger bright spots in Fig.~\ref{STM}(b)). After annealing to $T_{\rm A} \approx 370$~K (Fig.~\ref{STM}(c)) the density of the triangular-shaped depressions increases significantly accompanied by an increase in the size of the clusters, at the expense of the number of Fe adatoms. The same qualitative \ak{behavior} is found for Ca-doped samples (Figs.~\ref{STM}(d,e)). However, we focus our quantitative analysis on Fe on the pristine \BiSe~surface since the triangular-shaped depressions look similar to those of Ca dopants~\cite{Hor2009}.

In Fig.~\ref{STM}(f) the number of Fe adatoms and triangular-shaped depressions relative to the initial number of deposited Fe adatoms, as a function of $T_{\rm A}$, is depicted. The abrupt decrease in the number of Fe adatoms at $T_{\rm A} \approx 275$~K coincides with the emergence of the depression defects.  This strongly indicates that the observed triangular-shaped depressions are the signature of Fe atoms \ak{incorporated} into a subsurface layer of the \BiSe~substrate, upon thermal activation. In order to further exclude that the triangular-shaped depressions originate from thermally-induced defects of the host material, e.g. from Bi vacancies, we did the following control experiment: Fe was initially deposited at RT and subsequently cooled down for STM measurements (not shown) revealing a similar number of triangular-shaped depressions as in the case where cold
deposition and subsequent RT annealing were performed. The density of triangular-shaped depressions correlates with the amount of deposited Fe as shown for a higher initial coverage of 6\% ML in Fig.~\ref{STM}(e).  \ak{Moreover, the clean \BiSe~surface \jw{without the presence of Fe does not exhibit such defects regardless of applied temperature.}}

\begin{figure*}[tbp]
\includegraphics[width=0.80\textwidth]{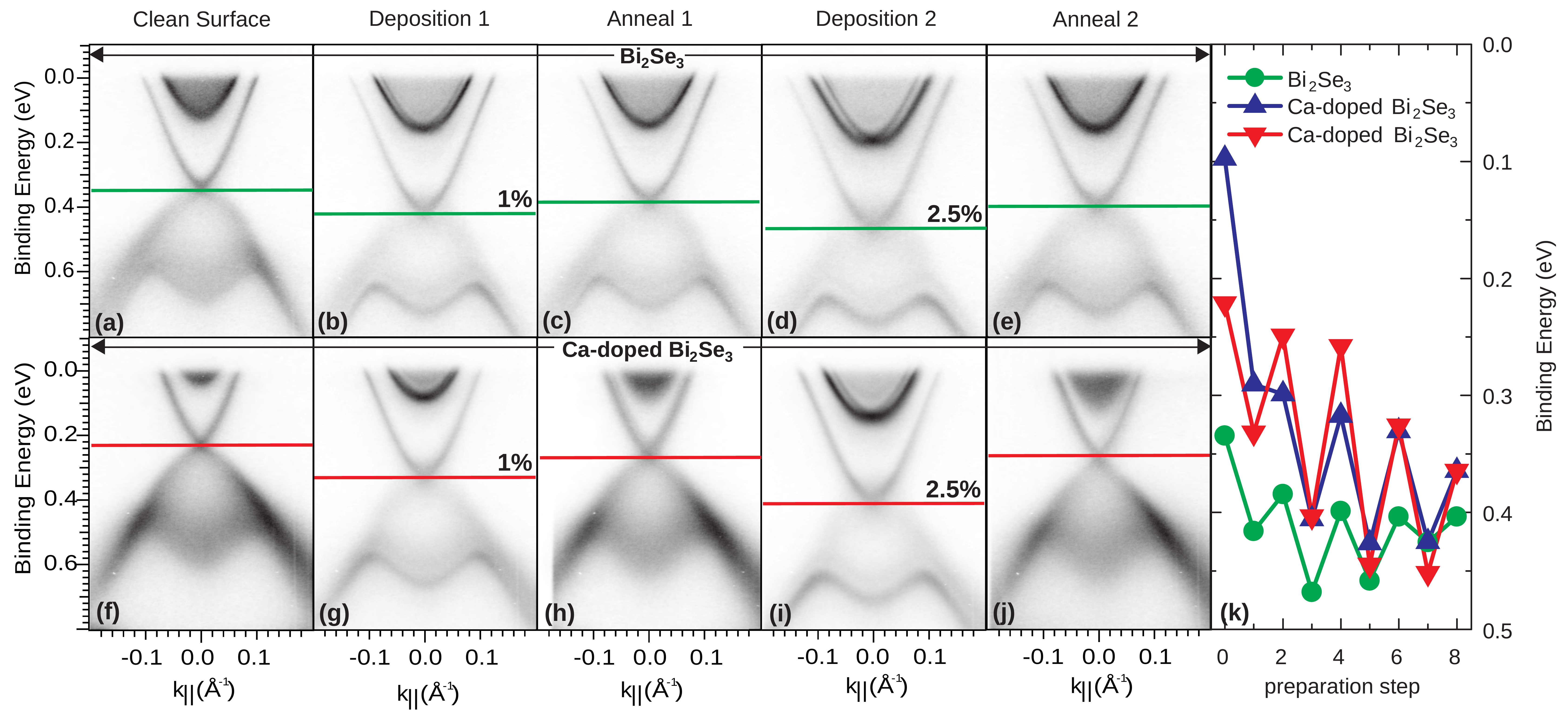}
\caption{\label{ARPES} Two series of ARPES measurements for pristine (a-e) and Ca-doped (f-j) \BiSe~samples \jw{after different Fe deposition and annealing steps}. (a), (f) Initial spectra after \textit{in-situ} cleaving. Subsequent cycles of cold deposition of Fe ($T \approx 150$K, \%~ML~Fe as indicated and annealing at $T_{\rm A} \approx 370$K followed. The DP extracted from the spectra is marked by a green and red horizontal line, respectively. (k) The resulting binding energy of the DP is plotted in dependence of the preparation step number for a pristine (circles) and two differently Ca-doped samples (up and down triangles). Preparation step number 0 indicates the freshly cleaved substrate without Fe, preparation step 1 (1\%), 3 (2.5\%), 5 (5\%), 7 (10\%), and 2, 4, 6, 8 subsequent annealing steps.}
\end{figure*}

Closer analysis of the STM topographs provides a deeper insight into the subsurface doping process. For $T_{\rm A} \approx 275$~K (Fig.~\ref{STM}(b)), only one type of triangular-shaped depressions is found. Samples that were annealed at higher temperatures exhibit triangular-shaped depressions with at least two different appearances in STM topographs (Fig.~\ref{STM}(c)). Figs.~\ref{STM2}\ak{(a,b)} exemplarily show high resolution STM images of two such defects and Figs.~\ref{STM2}(c-h) show bias\ak{-}dependent STM images of one of those defects. Defects with a strikingly similar STM appearance have been recently observed for Fe bulk-doped Bi$_2$Te$_3$~\cite{Okada2011}, Fe adatoms on Bi$_2$Te$_3$ after annealing~\cite{West2012} and for Fe bulk-doped \BiSe~\cite{Song2012}. Such defects were attributed to substitutional Fe atoms within Bi sites in the first quintuple layer (QL), i.e. in the 2$^{\mathrm{nd}}$ and 4$^{\mathrm{th}}$ subsurface layers.  We thus conclude the observed triangular depressions are substitutional Fe atoms residing in Bi sites. \jw{Moreover, as pointed out in Ref.~\cite{Song2012}, Fe atoms substituted into Bi sites in deeper QL are too far below the surface to be imaged by STM.} This might explain part of the \ak{discrepancy between the deposited Fe and \jw{the number of Fe substitutional defects counted after annealing} (Fig.~\ref{STM}(f))}. Besides, emerging clusters \ak{upon annealing}, which \ak{may contain both Bi and Fe}, may also account for this discrepancy.

To further investigate the impact of Fe doping \ak{on the TSS}, ARPES measurements were performed. In Fig.~\ref{ARPES}, two series of ARPES measurements are shown for pristine (a-e) and Ca-doped (f-j) \BiSe~substrates, at different stages of the preparation procedure. After \textit{in-situ} cleaving (a,f), subsequent iterations of cold Fe deposition ($T \approx 150$~K) and annealing to $T_{\rm A} \approx 370$~K were performed with no cleaving in between. The amount of subsequently deposited Fe was increased for each iteration (1\%, 2.5\%, 5\%, 10\%). The ARPES data after each cold deposition \ak{\textit{but before subsequent annealing}} (\ak{Fig.~\ref{ARPES} (b,d,g,i)}) reveals a downward shift of the DP \ak{as compared \jw{to its energetic position before Fe deposition.}} We can thus conclude that Fe adatoms are donors \ak{inducing} downwards band bending.  \ak{Such band bending leads to the} well known formation of quantum well states (QWS)~\cite{Bianchi2010,Bianchi2011,Wray2011,Benia2011,Valla2012,Bianchi2012} in both the conduction and the valence bands~\cite{Bianchi2011}, where the conduction band QWS shows a clear Rashba splitting.

Surprisingly, after annealing to temperatures \ak{coinciding with} STM \ak{observations of Fe diffusion} into the subsurface region \ak{(Fig.~\ref{STM},\ref{STM2})}, the DP shifts back \ak{toward its initial position} and the surface's electronic band structure is reverted \ak{close to} its initial state before Fe deposition (Fig.~\ref{ARPES}~(c,e,h,j)). \ak{Strikingly, after annealing,} the QWS are \ak{nearly destroyed}. In order to further analyse this \ak{behavior}, the binding energy of the DP after each preparation step of the subsequent iterations has been extracted from the \ak{ARPES} spectra and is plotted in Fig.~\ref{ARPES}(k) for the pristine and two Ca-doped substrates with different Ca concentrations. It reveals the afore mentioned oscillation of the DP binding energy after each preparation step, i.e. an increase in the DP binging energy for cold deposition of Fe, and a decrease after annealing the Fe covered sample. The gradual shifting of the DP toward higher binding energies is most likely the result of the well known aging effect of the \BiSe~surface~\cite{Benia2011}.  In order to exclude the influence of rest gas physisorption at low temperature that would also lead to a downward band bending, the following test has been performed. The very same procedure of Fe deposition has been followed but keeping the shutter of the Fe evaporator closed and no appreciable change in the band bending has been observed after this simulated deposition. Our result is contrary to a recent ARPES study that concludes surface electron doping for Fe deposition at RT and hole doping for cold deposited Fe~\cite{Scholz2012}. Taking into account the STM-observed thermally-induced subsurface diffusion of Fe adatoms (Fig.~\ref{STM}~(f)), our sequence of \ak{ARPES} experiments indicates, that substitutional Fe atoms are either neutral dopants or electron acceptors, while the Fe adatoms that are prepared by cold deposition are electron donors. In all of these cases the DP stays intact without the opening of a gap.



\begin{figure}[tbp]
\includegraphics[width=0.95\columnwidth]{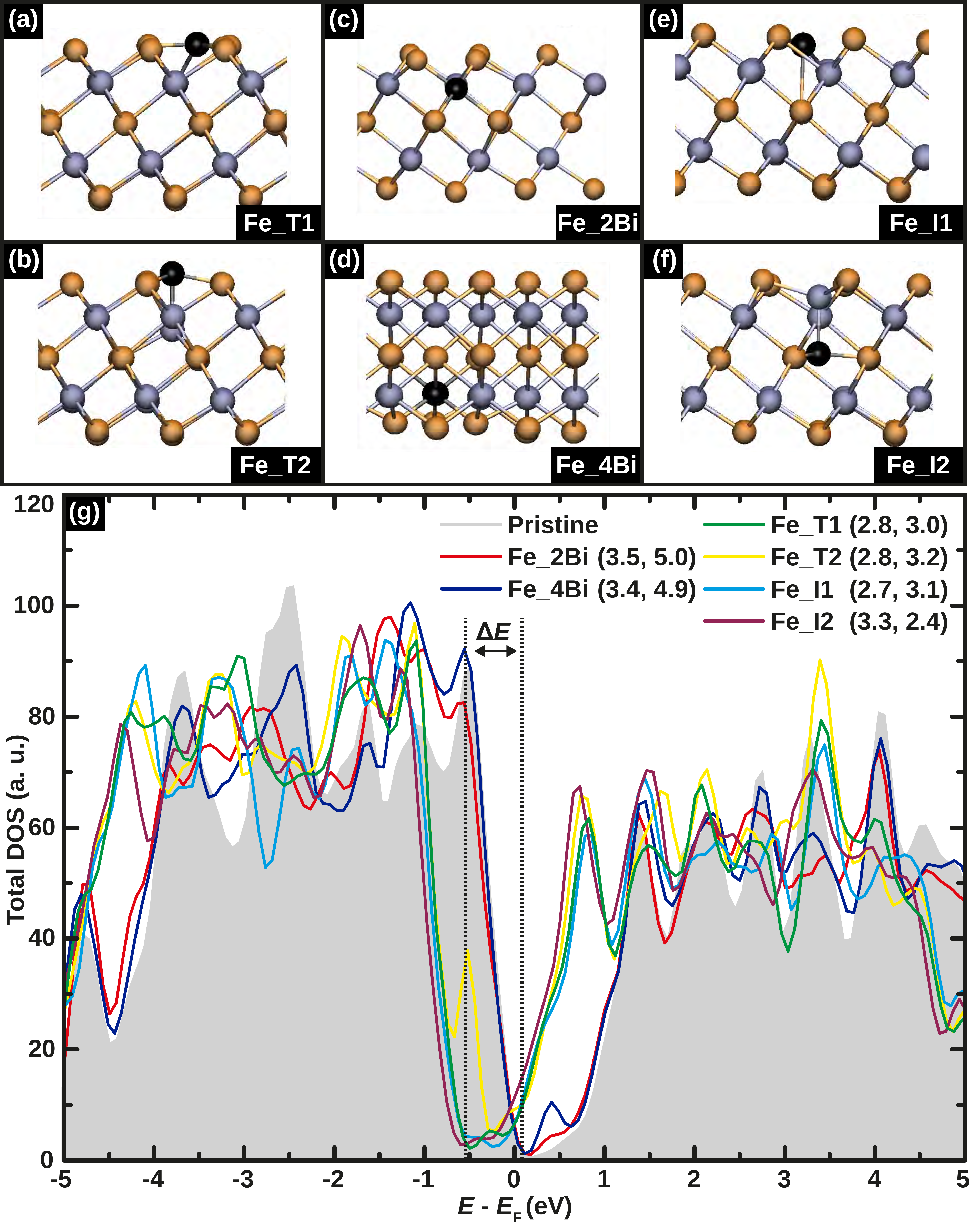}
\caption{\label{DFT} \jw{(a-f) Side views of ball and stick models of the different Fe doped \BiSe~systems considered in the \textit{ab-initio} calculations: (a,b) Fe on the two different hollow sites fcc (Fe\underline{ }T1) and hcp (Fe\underline{ }T2), (c,d) Fe substituting Bi atoms in the 2$^{\mathrm{nd}}$ (Fe\underline{ }2Bi) and 4$^{\mathrm{th}}$ (Fe\underline{ }4Bi) subsurface layer, and (e,f) Fe in two different interstitial sites (Fe\underline{ }I1, Fe\underline{ }I2) of the \BiSe~lattice. Fe is shown in black, Bi grey, and Se orange. (g) Calculated total DOS for the different considered sites. The resulting energy shift $\Delta E$ of the DOS due to the doping is indicated. The Fe $3d$ magnetic moments and the total magnetic moments ($m_{\rm Fe}$, $m_{\rm tot}$) for Fe at these different sites are given in the legend in units of $\mu_B$.}}
\end{figure}

In order to determine the electronic and magnetic properties of the incorporated Fe atoms\ak{,} we performed \textit{ab initio} calculations. \jw{As shown in the ball and stick models in Fig.~\ref{DFT}~(a-f), we considered pure \BiSe, as well as Fe on the two different hollow sites fcc (a) and hcp (b), Fe substituting Bi atoms in the 2$^{\mathrm{nd}}$ (c) and 4$^{\mathrm{th}}$ (d) subsurface layer, and Fe in two different interstitial sites (e,f) of the \BiSe~lattice.} The calculated total density of states (DOS) for these different cases is illustrated in \jw{Fig.~\ref{DFT}~(g)}. For pure \BiSe~(grey shaded)\ak{,} the onset of the valence band \ak{emerges at} energies below the Fermi energy $E_{\rm F}$ and the onset of the conduction band DOS for $E-E_{\rm F} > 0.75$~eV. The minimum and gradual increase in the DOS within the bulk band gap ($E-E_{\rm F} < 0.75$~eV) are due to the DP of the TSS. Compared to the DOS of the pure \BiSe, the Fe adatoms (green, yellow) or the Fe subsurface interstitials (\jw{cyan and purple}) cause a shift of the bulk band gap and DP to a higher binding energy by $\Delta E\approx 0.5$~eV. This \ak{reproduces the} electron doping behavior of the Fe adatoms, \ak{as observed by} ARPES. Moreover, Fe atoms \ak{in} Bi substitutional sites (red, blue) do not shift the bulk band gap and DP with respect to the pure \BiSe, indicating neutral doping for these impurities. Note here that the Fermi energy is always below the onset of the conduction band within the calculations, while in the experiment it is slightly shifted into the conduction band due to Se vacancy doping and aging effects. For a strong shift of the Fermi energy, the Fe atoms on Bi substitutional sites might \ak{become} electron acceptors. These conclusions are consistent with the above interpretation of the STM and ARPES data, confirming that the thermally-induced subsurface Fe doping results in a preferential occupation of Bi substitutional sites, as already suggested by STM.

Finally, we calculated the magnetic spin moments of the Fe atoms in the different lattice sites as indicated in the legend of \jw{Fig.~\ref{DFT}~(g)}. For all defects, we find considerable magnetic moments $2.7\mu_B<m_{\rm Fe}<3.5\mu_B$ in the Fe $d$ shell. $m_{\rm Fe}$ is largest for Fe substitutional defects, where Fe is nominally in a $3^+$ oxidation state. In contrast, Fe adatoms are clearly in a lower oxidation state as no Bi$^{3+}$ is missing. Thus, the reduced magnetic moments of Fe adatoms as compared to substitutional Fe atoms as well as the stronger tendency of Fe adatoms to donate further charge to the \BiSe~conduction band is well understandable. \ak{Fe substitutional atoms thus not only minimize the charge doping but also maximize the magnetic coupling to the \BiSe~host.}

\jw{For a magnetism-induced modification of the TSS by Fe, the coupling of the Fe $3d$ magnetic moments to the \BiSe~host is decisive.} It is therefore illustrative to analyze also the total spin magnetic moments \jw{$m_{\rm tot}$} of the supercell. While the total magnetic moments $m_{\rm tot}$ and $m_{\rm Fe}$ are rather similar in the adatom case, $m_{\rm tot}$ exceeds $m_{\rm Fe}$ by $\sim 1.5\mu_B$ in the case of Fe substitutionals. In the latter case a sizable magnetic moment is induced in the vicinity of Fe which indicates much stronger magnetic coupling between substitutional Fe atoms and the \BiSe~host than for adatoms.


In summary, it was demonstrated that, via low temperature deposition and moderate annealing, neutral magnetic moments can be introduced into the topmost layers of a prototypical TI, thus preserving the bulk electronic properties of the TI without causing significant band bending effects at the surface. In the investigated case of Fe on \BiSe, Fe acts as an electron donor when adsorbed on top of the surface at low temperature, but \ak{becomes} \jw{neutral or an acceptor} when substituted into a subsurface Bi site by thermally-activated diffusion. This allows for a study of the time-reversal symmetry breaking driven scattering of the TSS electrons by magnetic impurities.

We acknowledge financial support from the DFG via SFB~668, by the ERC Advanced Grant "FURORE," by the city of Hamburg via the cluster of excellence "Nanospintronics," by the Danish Council for Independent Research, and the Danish National Research Foundation.


\begin{thebibliography}{10}
\newcommand{\enquote}[1]{`#1'}

\bibitem{Hasan2010}
M.~Hasan, C.~Kane, \textit{Rev. Mod. Phys.}, \textbf{82}, 3045 (2010).

\bibitem{Qi2011}
X.-L. Qi, S.-C. Zhang, \textit{Rev. Mod. Phys.}, \textbf{83}, 1057 (2011).

\bibitem{Liu2009}
Q.~Liu, \textit{et~al.}, \textit{Phys. Rev. Lett.}, \textbf{102}, 156603
  (2009).

\bibitem{Biswas2010}
R.~Biswas, A.~Balatsky, \textit{Phys. Rev. B}, \textbf{81}, 233405 (2010).

\bibitem{Chen2010}
Y.~L. Chen, \textit{et~al.}, \textit{Science}, \textbf{329}, 659 (2010).

\bibitem{Wray2011}
L.~A. Wray, \textit{et~al.}, \textit{Nature Phys.}, \textbf{7}, 32 (2011).

\bibitem{Liu2012}
M.~Liu, \textit{et~al.}, \textit{Phys. Rev. Lett.}, \textbf{108}, 036805
  (2012).

\bibitem{Xu2012}
S.-Y. Xu, \textit{et~al.}, \textit{Nature Phys.}, \textbf{8}, 616 (2012).

\bibitem{Hor2010}
Y.~S. Hor, \textit{et~al.}, \textit{Phys. Rev. Lett.}, \textbf{104}, 057001
  (2010).

\bibitem{Vobornik2011}
I.~Vobornik, \textit{et~al.}, \textit{Nano Lett.}, \textbf{11}, 4079 (2011).

\bibitem{Valla2012}
T.~Valla, \textit{et~al.}, \textit{Phys. Rev. Lett.}, \textbf{108}, 117601
  (2012).

\bibitem{Scholz2012}
M.~Scholz, \textit{et~al.}, \textit{Phys. Rev. Lett.}, \textbf{108}, 256810
  (2012).

\bibitem{Bianchi2011}
M.~Bianchi, \textit{et~al.}, \textit{Phys. Rev. Lett.}, \textbf{107}, 086802
  (2011).

\bibitem{Bianchi2010}
M.~Bianchi, \textit{et~al.}, \textit{Nature Commun.}, \textbf{1}, 128 (2010).

\bibitem{King2011}
P.~King, \textit{et~al.}, \textit{Phys. Rev. Lett.}, \textbf{107}, 096802
  (2011).

\bibitem{Beidenkopf2011}
H.~Beidenkopf, \textit{et~al.}, \textit{Nature Phys.}, \textbf{7}, 939 (2011).

\bibitem{Eremeev2012}
S.~V. Eremeev, \textit{et~al.}, \textit{Nature Commun.}, \textbf{3}, 635
  (2012).

\bibitem{Hoffmann2004}
S.~Hoffmann, \textit{et~al.}, \textit{Nucl. Instrum. Methods Phys. Res. Sec.
  A}, \textbf{523}, 441  (2004).

\bibitem{Bianchi2012}
M.~Bianchi, \textit{et~al.}, \textit{ACS Nano}, \textbf{6}, 7009 (2012).

\bibitem{Honolka2012}
J.~Honolka, \textit{et~al.}, \textit{Phys. Rev. Lett.}, \textbf{108}, 256811
  (2012).

\bibitem{PhysRevB.47.558}
G.~Kresse, J.~Hafner, \textit{Phys. Rev. B}, \textbf{47}, 558 (1993).

\bibitem{PhysRevB.54.11169}
G.~Kresse, J.~Furthm\"uller, \textit{Phys. Rev. B}, \textbf{54}, 11169 (1996).

\bibitem{PhysRevLett.77.3865}
J.~P. Perdew, K.~Burke, M.~Ernzerhof, \textit{Phys. Rev. Lett.}, \textbf{77},
  3865 (1996).

\bibitem{PhysRevB.50.17953}
P.~E. Bl\"ochl, \textit{Phys. Rev. B}, \textbf{50}, 17953 (1994).

\bibitem{PhysRevB.59.1758}
G.~Kresse, D.~Joubert, \textit{Phys. Rev. B}, \textbf{59}, 1758 (1999).

\bibitem{Hor2009}
Y.~Hor, \textit{et~al.}, \textit{Phys. Rev. B}, \textbf{79}, 195208 (2009).

\bibitem{Okada2011}
Y.~Okada, \textit{et~al.}, \textit{Phys. Rev. Lett.}, \textbf{106}, 206805
  (2011).

\bibitem{West2012}
D.~West, \textit{et~al.}, \textit{Phys. Rev. B}, \textbf{85}, 081305(R) (2012).

\bibitem{Song2012}
C.-L. Song, \textit{et~al.}, \textit{Phys. Rev. B}, \textbf{86}, 045441 (2012).

\bibitem{Benia2011}
H.~Benia, C.~Lin, K.~Kern, C.~Ast, \textit{Phys. Rev. Lett.}, \textbf{107},
  177602 (2011).

\end{thebibliography}

\end{document}